\documentstyle[eqsecnum,aps,prd]{revtex}
\begin{document}
\newcommand{\gsim}{\mbox{\raisebox{-1.0ex}{$\stackrel{\textstyle >}
{\textstyle \sim}$ }}}
\newcommand{\lsim}{\mbox{\raisebox{-1.0ex}{$\stackrel{\textstyle <}
{\textstyle \sim}$ }}}
\newcommand{\bfx}{{\bf x}}
\newcommand{\bfy}{{\bf y}}
\newcommand{\bfr}{{\bf r}}
\newcommand{\bfk}{{\bf k}}
\newcommand{\bkp}{{\bf k'}}
\newcommand{\order}{{\cal O}}
\newcommand{\beq}{\begin{equation}}
\newcommand{\eeq}{\end{equation}}
\newcommand{\beqa}{\begin{eqnarray}}
\newcommand{\eeqa}{\end{eqnarray}}
\newcommand{\lmk}{\left(}
\newcommand{\rmk}{\right)}
\newcommand{\lkk}{\left[}
\newcommand{\rkk}{\right]}
\newcommand{\lnk}{\left\{}
\newcommand{\rnk}{\right\}}
\newcommand{\call}{{\cal L}}
\newcommand{\calh}{{\cal H}}
\newcommand{\phic}{\phi_c}
\newcommand{\pphi}{p_{\phi_c}}
\thispagestyle{empty}
\thispagestyle{empty}
{\baselineskip0pt
\leftline{\large\baselineskip16pt\sl\vbox to0pt{\hbox{Department of Physics}
               \hbox{Kyoto University}\vss}}
\rightline{\large\baselineskip16pt\rm\vbox to20pt{\hbox{KUNS-1327}
           \hbox{OCHA-PP-56}
           \hbox{YITP/U-95-24}
               \hbox{\today}
\vss}}%
}
\vskip15mm
\begin{center}
{\large\bf Dynamics of Subcritical Bubbles in First Order
Phase Transition}
\end{center}
\begin{center}
{\large Tetsuya Shiromizu} \\
\sl{Department of Physics, Kyoto University, Kyoto 606-01, Japan}
\end{center}
\begin{center}
{\large Masahiro Morikawa} \\
\sl{Department of Physics, Ochanomizu University, Tokyo 112, Japan}
\end{center}
\begin{center}
{\large Jun'ichi Yokoyama} \\
\sl{Uji Research Center, Yukawa Institute for Theoretical Physics \\
Kyoto University, Uji, Kyoto 611, Japan}
\end{center}
\begin{center}
{\it to be published in Progress of Theoretical Physics}
\end{center}
\begin{abstract}
We derivate the Langevin and the Fokker-Planck equations
for the radius of $O(3)$-symmetric subcritical bubbles
as a phenomenological model to treat thermal fluctuation.
The effect of  thermal noise on subcritical bubbles
is examined. We find that the fluctuation-dissipation relation holds
and that in the high temperature phase
the system settles down rapidly to the
thermal equilibrium state even if it was in a nonequilibrium
state initially.
We then estimate the typical size of subcritical bubbles as well as
the amplitude of fluctuations on that scale.
We also discuss their implication to
the electroweak phase transition.
\end{abstract}
\vskip1cm



\section{Introduction}

The dynamics of the electroweak phase transition is important
in the light of electroweak baryogenesis \cite{ctm}.
For the scenario to be successful the phase transition must be of
first order with supercooling in order to attain a nonequilibrium
state necessary for baryogenesis \cite{sak}.
Though the effective potential of the Higgs
field obtained by perturbation possesses a barrier between  false
and true vacua at the critical temperature \cite{ew}, it has  not been
certain if the phase transition accompanies supercooling.
The purpose of this paper is to negatively confirm this.
If the thermal fluctuation around the symmetric vacuum is too large,
the phase transition proceeds without supercooling and no
baryogenesis is expected.

In the last five years, some attempts have been made by several
different methods \cite{ew} -- \cite{shiromizu}.
For example, Gleiser and Kolb \cite{gl1}\cite{gl4} have discussed that
the conventional picture of the first-order phase transition
through nucleation of critical bubbles is applicable in the minimal
standard model only in the case of relatively light Higgs mass,
$m_H \leq 70$GeV, and that otherwise subcritical fluctuations plays a
dominant role to realize emulsion of false and true vacua even above
the critical temperature.  In their analysis, however, it has been
assumed that the typical scale of subcritical bubbles is given by the
correlation length of the Higgs field.
Meanwhile, as a way to understand the dynamics of quantum fields in a
finite-temperature but nonequilibrium situation, Gleiser and Ramos
\cite{gl4} have derived the Langevin equation for a scalar field
extending Morikawa's method \cite{mori}.
The equation has also been numerically analyzed assigning white noise
on a lattice, and it has been concluded that the sufficient phase
mixing happen in any experimentally-allowed range of Higgs mass,
where again the
lattice spacing has been taken to be comparable to the correlation
length\cite{gl5}.
Because the amplitudes of thermal fluctuation changes
drastically depending on the length scale, it is important to
determine physically the typical
length scale of fluctuations which dominate
the dynamics of the phase transition.
As a first step along this line, in a previous paper \cite{shiromizu}
we have obtained the typical scale of $O(3)$-symmetric
subcritical bubbles by
constructing the effective Hamiltonian for their radius
and taking a thermal average.  We thereby concluded that the
electroweak phase transition is dominated by subcritical bubbles with
any experimentally-allowed value of Higgs mass.
Unfortunately, however, we cannot deny the fact that our approach was
also too phenomenological to be viable from fundamental points of
view.

In the present paper, applying nonequilibrium statistical field
theory we derive the Langevin and the Fokker-Planck
equations for the Higgs field
to clarify the detailed structure of the electroweak
phase transition. Starting with the effective action we first
derive these equations for generic field configurations.
We then adopt a Gaussian ansatz for field configuration
to model a subcritical bubble and yield the effective Langevin
and Fokker-Planck equations for its radius. This is a kind of
variational approach which is adopted because, unlike the
critical bubble, these subcritical bubbles do not constitute
a solution of field equations but should merely be regarded
as a model of field configurations.
As a result our previous approach will be justified under some
conditions.
Since we construct the Langevin equation in the
perturbation theory, the effect from the environment is automatically
taken into account. We need to calculate some loop corrections to obtain
the friction term. The friction
term will be estimated in the quasi-adiabatic approximation, while the noise
term comes from the imaginary part of the effective action.
The Fokker Planck equation can
be derived directly from the Langevin equation and then we find that
its static solution is of the form which we assumed as a
probability amplitude in the previous
paper.

The rest of the present  paper is organized as follows.
In Sec. II, we review the
non-equilibrium quantum field theory and discuss the statistical aspects
of the  theory. In Sec.III, we
derive the Langevin and Fokker-Planck equations for the radius of
subcritical bubbles. We examine the effect of dissipation and thermal noise
on the electroweak phase transition. Section IV is devoted to summary
and discussion.
Throughout the paper we use the units  $c=\hbar=1$.


\section{Effective Lagrangian of Higgs fields in the Thermal Bath}

\subsection{Basics of the non-equilibrium quantum field theory}

In order to derive the Langevin equation and the Fokker-Planck
equation for the subcritical bubbles, we use the non-equilibrium
quantum field theory.  This subsection is a brief review of its
basic technique
\cite{cqft}.

The ordinary quantum field theory, which mainly deals with
transition amplitudes in particle reactions,  is not useful
when we study statistical dynamics of macroscopic objects like bubbles.
This is because we need the temporal evolution of some kind of
classical order parameters with definite initial condition and not
simply the transition amplitude of particle reactions with fixed
initial and final conditions.  The most appropriate extension of the
field theory to deal with these issues is to generalize
the time contour of integration to the closed form.  More precisely,
the time integration contour is generalized so that it runs from
minus infinity to plus infinity and then back to the minus infinity
again.   This formalism, often called as in-in formalism of quantum
field theory,
yields various quantum averages of operators evaluated in the
in-state without specifying out state. On the other hand the
ordinary quantum field theory, often called as in-out formalism
of quantum field theory, yields quantum averages of operators
evaluated with an in-state at one end and an out-state at the other.

The partition function in the in-in formalism for a real scalar field is
defined to be
%
\begin{eqnarray}
Z[J]&\equiv& Tr\lkk T\lmk{\rm exp}\lkk i\int_c { J \varphi }\rkk
\rmk\rho\rkk
\nonumber \\
&=&Tr\lkk T_+\lmk{\rm exp}\lkk i\ \int { J_+ \varphi_+ }\rkk\rmk T_-
\lmk{\rm exp}
\lkk -i\ \int { J_- \varphi_- }\rkk\rmk\rho\rkk
\end{eqnarray}
%
where the suffix $c$ in the integral means that the time integration
contour runs from minus infinity to plus infinity and then back to
the minus infinity again.   All the field quantity is defined on this
closed time contour.  In the above, $X_+$ represents a field
component $X$ which is restricted on the forward branch
($-\infty$ to $+\infty$) of the time contour and $X_-$ stands for
that restricted on the backward
branch ($+\infty$ to $-\infty$).  In the rest of this paper we often
use the following notation:
$X_{\Delta}=X_+ - X_-$ and $X_C=(X_+ + X_-)/2$.
The symbol $T$ designates the operator ordering with respect to
this closed time contour, and accordingly $T_+$ designates the
ordinary time ordering and $T_-$ the anti-time ordering.
Here $J$ means the external field.  Although $J_+$ and $J_-$are the
same actually, we regard that they are
different from each other for technical reasons
and we only set $J_+=J_-$ at the end
of calculations.
The symbol $\rho$ represents the initial density matrix, and the
field $\varphi(x)$ is in the Heisenberg picture.

In the interaction picture, this partition function becomes
%
\begin{eqnarray}
Z[J]&=&Tr\lkk T\lmk{\rm exp}\lkk i\int_c
{ J \varphi +i\int_c {V[\varphi ]}}\rkk\rho \rmk\rkk
\nonumber \\
    &=&{\rm exp}\Bigl(-i\int_c {V\lkk{\delta  \over {i\delta  J}}\rkk}\Bigr)
Tr\lnk T\lkk{\rm exp}\lmk i\int_c { J \varphi }\rmk\rho \rkk\rnk \nonumber\\
    &=&{\rm exp}\Bigl(-i\int_c {V\lkk{\delta  \over {i\delta  J}}\rkk} \Bigr)
{\rm exp}\Bigl[-{i \over 2}\int_c \int_c { J(x) G_0(x,y)J(y)}\Bigr].
\end{eqnarray}
%
where we have used the Wick theorem which holds not only in the vacuum
state but also in the thermal state with $\rho=\exp(-H/T)$.
In the latter the above propagator is, in the momentum
representation,
%
\begin{eqnarray}
G_0(p)&=&
\left( \begin{array}{cc}
G_F(p) & G_+(p) \\
G_-(p) & G_{\bar F}(p)
\end{array} \right)
\nonumber\\
&=&
\left( \begin{array}{cc}
{1 \over p^2-m^2+i \epsilon}-2\pi i n({\bf p})\delta(p^2-m^2) &
-2\pi i (\theta(p_0)+n({\bf p}))\delta(p^2-m^2)\\
-2\pi i (\theta(-p_0)+n({\bf p}))\delta(p^2-m^2) &
{-1 \over p^2-m^2-i \epsilon}-2\pi i n({\bf p})\delta(p^2-m^2),
\end{array} \right)
\end{eqnarray}
%
where $n(\bf p)$ is the thermal distribution function:
$n({\bf p})=[\exp(\omega({\bf p})/T)-1]^{-1}$ and $\omega=
\sqrt{{\bf p}^2+m^2}$.

By the Legendre transformation of the partition function,
we obtain the generalized effective action, or generating
functional of the vertex functions:
%
\begin{eqnarray}
\phi (x)&\equiv& {{\delta \ln Z} \over {i\delta  J(x)}}, \nonumber \\
\Gamma [ \phi ]&\equiv& -i\ln Z[ J]-\int_c { J \phi }.
\end{eqnarray}
%
The equality
%
\begin{equation}
{{\delta \Gamma [\phi ]} \over {\delta \phi (x)}}=-J(x)
\end{equation}
%
immediately follows as in the in-out formalism and this form is
often used as a generalized classical equation of motion for the
variable $\phi(x)$.  However, this
effective action has an imaginary part.
For example if we parameterize the kernel of
two-point part of
$\Gamma$ as \cite{cqft}
%
\begin{equation}
{\hat\Gamma}^{(2)}(x,y)=
\left( {\matrix{D+iB&i(B-A)\cr
i(B+A)&-D+iB\cr
}} \right),
\end{equation}
%
the imaginary part of ${\Gamma}^{(2)}$ is
%
\begin{equation}
{\rm Im} {\Gamma}^{(2)} [\phi _c,\phi _\Delta ]={1 \over 2}\int\!\!\!\int
{\phi _\Delta (x)B(x-x')\phi _\Delta (x')}.
\end{equation}
%
We can rewrite this expression by introducing a real auxiliary
field $\xi(x)$ \cite{mori}:
%
\begin{equation}
{\rm exp}\lmk i\Gamma [\phi ]\rmk =\int {[d\xi ]}P[\xi ]
{\rm exp}\lkk i {\rm Re} \Gamma +\int {i\xi \phi _\Delta }\rkk
\end{equation}
%
where,
%
\begin{equation}
P[\xi ]={\rm exp}\lkk-{1 \over 2}\int \!\! \int {\xi B^{-1}\xi }\rkk
\end{equation}
%
is a normalizable positive definite statistical weight for the fields $\xi(x)$.
If we apply the variational principle for the exponent of the
integrand of eq.(2.8), we obtain an equation of motion for
$\varphi_c(x)$ as
%
\begin{eqnarray}
-J_C&=& \left({{\delta  {\rm Re}\Gamma +\int \xi \phi_{\Delta} }
	\over {\delta  \phi_{\Delta} (x)}} \right)_{\phi_{\Delta}=0} \nonumber \\
&=& \Box \phi_c + V'(\phi_c)+
\int dx'A(x-x') \phi_c (x') -\xi(x).
\end{eqnarray}
%
This is a Langevin type stochastic differential equation with
a nonlocal kernel $A(x-x')$.
If the time scale of change in $\phi_c(x)$ is small compared with that of
radiative corrections and $\phi_c(x)$ is nearly homogeneous in space,
this term reduces to the familiar friction term:
%
\begin{equation}
\Box \phi_c (x)+ V'_{\rm eff}(\phi_c(x))+
\eta \dot \phi_c (x)  =\xi(x),
\end{equation}
%
with
%
\begin{equation}
\eta=-i{\rm lim}_{k_{\mu} \to 0}{\partial A(k) \over \partial k^0},
\end{equation}
%
where $V_{\rm eff}$ is the effective potential and $A(k)$ is the
Fourier transform of $A$.

If we define the statistical average  as
%
\begin{equation}
\left\langle {...} \right\rangle _\xi \equiv \int {[d\xi ]}P[\xi ]...,
\end{equation}
%
we then obtain
%
\begin{equation}
\left\langle {\xi (x)\xi (x')} \right\rangle _\xi =B(x-x').
\end{equation}
%
Thus we can construct consistent statistical field theory in
the in-in formalism of quantum field theory.

In the actual application of this formalism to the subcritical
bubbles, we need to calculate the dissipative and diffusive
kernels $A$ and $B$ at finite temperature.
This is the subject of the next two subsections.

\subsection{The fluctuation-dissipation relation and
the stationary distribution for Higgs fields}

For simplicity, we consider the following Lagrangian of a singlet
Higgs field $\varphi$.
%
\begin{equation}
{\cal L}=\frac{1}{2}({\partial}_{\mu} \varphi)^2 -\frac{1}{2}m^2
{\varphi}^2 -\frac{1}{4!}\lambda {\varphi}^4+ i {\overline {\psi}}
{\gamma}^{\mu} {\partial}_{\mu} \psi
-f \varphi  {\overline {\psi}} \psi.
\end{equation}
%
Strictly, we must include interactions with $Z$, $W$-bosons
and all quarks for the detail of the electroweak phase transition.
In particular, contributions of  $Z$ and $W$ boson are crucial to
yield the cubic term in the effective potential with
one-loop corrections.
However, the
essence of non-equilibrium phase transition can be fully
obtained by the above
simple model as we will see soon.
According to the calculation by Morikawa\cite{mori} and
Gleiser et al\cite{gl4}, the
effective action becomes, up to one-loop corrections,
%
\begin{eqnarray}
{\Gamma}[{\phi}_c,{\phi}_{\Delta}] & = & {\int}d^4x\Bigr\lbrace
{\phi_{\Delta}}(x) [ -\Box -V(t)]
{\phi}_c(x)-\frac{\lambda}{4!}(4{\phi}_{\Delta}(x){\phi}^3_c(x)
+{\phi}_{\Delta}^3(x){\phi}_c(x)) \Bigr\rbrace \nonumber \\
& -2 & {\int}d^4xd^4x'A_1(x-x'){\phi}_{\Delta}(x){\phi}_c(x') -
{\int}d^4xd^4x' A_2(x-x')[{\phi}_{\Delta}(x){\phi}_c(x)
{\phi}_c^2(x')+\frac{1}{4}{\phi}_{\Delta}{\phi}_c(x){\phi}_{\Delta}^2
(x')]
\nonumber \\
& + & \frac{i}{2}{\int}d^4xd^4x' \Bigl[ B_1(x-x')
{\phi}_{\Delta}(x){\phi}_{\Delta}(x')+
B_2(x-x'){\phi}_{\Delta}(x)
{\phi}_{\Delta}(x'){\phi}_c(x){\phi}_c(x')\Bigr],
\end{eqnarray}
%
where
%
\begin{equation}
V(t):=m^2+\frac{{\lambda}}{2}{\int}\frac{d^3k}{(2 \pi)^3}
\frac{1+2n(\omega)}{2\omega({\bf k})},
\end{equation}
%
%
\begin{equation}
A_1(x-x'):=f^2{\rm Re}[-iS^F_{\alpha \beta}(x-x')S_F^{ \beta \alpha}
(x'-x)]\theta (t-t'),
\end{equation}
%
%
\begin{equation}
A_2(x-x'):=\frac{{\lambda}^2}{2}{\int}\frac{d^3k}{(2 \pi)^3}
e^{i{\bf k} \cdot ({\bf x}-{\bf x}')} {\int}\frac{d^3q}{(2 \pi)^3}
{\rm Im}[G_F({\bf q}, t-t')G_F({\bf q}-{\bf k}, t-t')]{\theta}(t-t'),
\end{equation}
%
%
%
%
\begin{equation}
B_1(x-x'):=f^2{\rm Im}[-iS^F_{\alpha \beta}(x-x')S_F^{ \beta \alpha}
(x'-x)],
\end{equation}
%
%
\begin{equation}
B_2(x-x'):= \frac{{\lambda}^2}{2}{\int}
\frac{d^3k}{(2 \pi)^3}e^{i{\bf k} \cdot ({\bf x}-{\bf x}')}
{\rm Re}{\int}\frac{d^3q}{(2 \pi)^3}[G_F({\bf q}, t-t')G_F
({\bf q}-{\bf k}, t-t')].
\end{equation}
%
%
%
Here $S_{\alpha \beta}^F(x-x')$ is the thermal Green's function
of the fermion defined by
%
\begin{eqnarray}
S_F(x-x') & := & -i Tr [ T_p({\psi}(x){\overline {\psi}}(x'))
\rho] \nonumber \\
& = & {\int}\frac{d^4p}{(2 \pi)^4}
\Bigl[\frac{1}{{\gamma}^{\mu}p_{\mu}+i \epsilon}
+ 2 \pi i {\gamma}^{\mu} p_{\mu}f({\bf p}) \delta (p^2) \Bigr]e^{-ip(x-x')},
\end{eqnarray}
%
where $f({\bf p})=({\rm exp}(|{\bf p}|/T)+1)^{-1}
$\footnote{Here we use a {\it massless} propagator as an
approximation. In fact once the phase transition starts, the top quark
has a space-time dependent mass due to the variation of the Higgs
field.  The maximum value of the top mass, however, remains as small
as $f \phi_0 \sim 50$GeV at the critical temperature $T_c \simeq
93$GeV.  We may therefore conclude that the result would not change
significantly even if we used the  more complicated propagator with
nontrivial space-time dependence or simply a massive propagator.}

We can rewrite this expression by introducing two real auxiliary
fields ${\xi}_1(x)$ and ${\xi}_2(x)$:
%
\begin{equation}
{\rm exp}(i\Gamma [\varphi ])=\int {[d{\xi}_1 ][d{\xi}_2] }P_1[{\xi}_1]
P_2[{\xi}_2]
{\rm exp}\Bigl[i {\rm Re} \Gamma
+i \int ( {{\xi}_1 \phi _\Delta +{\xi}_2 {\phi}_c
{\phi}_{\Delta} )}\Bigr],
\end{equation}
%
where
%
\begin{equation}
P_i[\xi ]={\rm exp}\lkk-{1 \over 2}\int \!\! \int
{{\xi}_i B_i^{-1}{\xi}_i }\rkk.
\end{equation}
%
Applying the variational principle on eq. (2.23) as in the
previous subsection, we obtain the Langevin equation:
%
\begin{eqnarray}
\Box \phi_c (x) & + & V'_{\rm eff}(\phi_c(x))
+2{\int}d^3x'{\int}^t_{- \infty} dt'{\phi}_c(x')A_1(x-x')
+ {\phi}_c(x){\int}d^3x'{\int}^t_{- \infty} dt'{\phi}_c^2(x')A_2(x-x')
\nonumber \\
&  = &
{\xi}_1(x)+ {\phi}_c(x) \xi_2(x).
\end{eqnarray}
%
Assuming that $\phi(x)$ is nearly
homogeneous and changes slowly in time, we obtain
%
\begin{equation}
\Box \phi_c (x)+ V'_{\rm eff}(\phi_c(x))+{\eta}_1 \dot {\phi}_c(x)
+\eta_2  {\phi}_c^2(x) \dot \phi_c (x)   ={\xi}_1(x)+{\phi}_c(x) \xi_2(x)
\end{equation}
%
with
%
\begin{equation}
{\eta}_1=2{\int}^{\infty}_0dt{\int}d^3{\bf x}A_1({\bf x},t)t ,
\end{equation}
%
%
\begin{equation}
{\eta}_2=2{\int}^{\infty}_0dt{\int}d^3{\bf x}A_2({\bf x},t)t ,
\end{equation}
%
%
%
and
%
\begin{equation}
\left\langle {{\xi}_i (x){\xi}_i (x')} \right\rangle _\xi =B_i(x-x').
\end{equation}
%
%
%
For the contribution from {\it only} the
self-coupling, Gleiser and Ramos\cite{gl4}
have obtained the result
$ {\eta}_2=\frac{96}{\pi T}{\ln {(\frac{T}{m_T})}}$ for
the friction coefficient and
%
\begin{equation}
\langle {\xi}_2(x) {\xi}_2(x') \rangle = 2T{\eta}_2{\delta}({\bf x}-
{\bf x}'){\delta}(t-t'),
\end{equation}
%
in the high temperature limit. However, for $\lambda \sim 10^{-2}$
(Higgs mass $m_H \sim 60$GeV) the correlation time of
the noise is $\Delta t_{\rm noise} \sim ({\rm Decay}~{\rm width})^{-1} \sim
1536 \pi (\lambda^2 T)^{-1} \sim 10^{7}T^{-1}$ which is much larger than
the typical scale $\sim T^{-1}$. Hence, the above approximation
is not guaranteed in \cite{gl4}.

In our present analysis, on the other hand, the most dominant contribution
comes from the Yukawa interaction with $1 \gsim f \gg \lambda$,
so that terms proportional to $\lambda^2$ are negligible and it
suffices to consider the following
Langevin equation:
%
\begin{equation}
\Box {\phi}_c +V'_{\rm eff}({\phi}_c) +{\eta}_1{\dot {\phi}}_c
= {\xi}_1.
\end{equation}
%
To obtain the friction term we must prepare the full
propagator of fermion. Up to one-loop order, the renormalized Green's
function becomes
%
\begin{eqnarray}
S_F({\bf p},t) & = & \frac{e^{-{\Gamma}_f|t|}}{2{\omega}_p}
\Bigl[ \Bigl(-{\gamma}^0{\omega}_p{\epsilon}(t)+\vec{\gamma} \cdot
\vec{p} \Bigr) f(-{\omega}_p+i{\Gamma}_f) e^{-i{\omega}_p|t|}  \nonumber \\
& - & \Bigl({\gamma}^0{\omega}_p{\epsilon}(t)+\vec{\gamma} \cdot
\vec{p} \Bigr) f({\omega}_p+i{\Gamma}_f) e^{i{\omega}_p|t|} \Bigr],
\end{eqnarray}
%
where ${\Gamma}_f$ is the decay width given by
%
\begin{equation}
{\Gamma}_f({\omega}_p)=-{\gamma}^0{\rm Im}(\Sigma)|_{p^2=0}
\simeq \frac{f^2}{8 \pi} T
\end{equation}
%
in the high temperature limit.
The friction coefficient then becomes
%
\begin{eqnarray}
{\eta}_1 & = &  2f^2{\rm Re}\Bigl[ i{\int}^{\infty}_0dt
{\int}\frac{d^3{\bf k}}{(2 \pi)^3}
S_{\alpha \beta}({\bf k},t) S^{\beta \alpha}({\bf k},-t)
\Bigr] t \nonumber \\
 & =& f^2 {\rm sin}(\beta {\Gamma}_f)
{\int}^{\infty}_0dte^{-2{\Gamma}_ft}t  {\int}
\frac{d^3{\bf k}}{(2 \pi)^3} \frac{1}{
{\rm cosh}(\beta {\omega}_k)+{\rm cos}(\beta {\Gamma}_f)} \nonumber \\
& = & \frac{f^2{\rm sin}(\beta {\Gamma}_f)}{8{\pi}^2{\Gamma}_f^2}
{\int}^{\infty}_0\frac{d \omega_k \omega_k^2}
{{\rm cosh}(\beta {\omega}_k)+{\rm cos}(\beta {\Gamma}_f)} .
\end{eqnarray}
%
Here we are using the approximation that the external momentum is
zero. In the case of Yukawa interaction with top quark, $f \sim 1$ and then
$\beta \Gamma_f \sim 0.04$. Thus, we can expand the last expression
by $\beta \Gamma_f$
and to the lowest order we find
%
\begin{equation}
\eta_1 \simeq \frac{f^2}{8{\pi}^2{\Gamma}_f^2}
\beta {\Gamma}_f {\int}^{\infty}_0\frac{d \omega_k \omega_k^2}
{{\rm cosh}(\beta {\omega}_k)+1}
=\frac{\pi}{3}T.
\end{equation}
%
Taking the same approximation used in the derivation of eq.(2.34),
the autocorrelation of noise becomes
%
\begin{eqnarray}
\langle {\xi}_1(x) {\xi}_1(x') \rangle & = & B_1(x-x') \nonumber \\
& = & -f^2{\rm Im}\Bigl[ i {\int}\frac{d^3{\bf k}}{(2 \pi)^3}
S_{\alpha \beta}({\bf k},t-t') S^{\beta \alpha}({\bf k},t'-t)
\Bigr] {\delta}^3({\bf x}-{\bf x}') \nonumber \\
& = & f^2{\int}\frac{d^3{\bf k}}{(2 \pi)^3}
\frac{e^{-2{\Gamma}_f|t-t'|}}{4[{\rm cosh}(\beta {\omega}_k)+{\rm cos}
(\beta \Gamma_f)]^2}
\Bigl\lbrace 2{\rm cos}(\beta \Gamma_f)
[{\rm cosh}(\beta {\omega}_k)+{\rm cos}(\beta \Gamma_f) ] \nonumber \\
& - & 2 {\rm sin}(\beta \omega_k){\rm sin}(\beta \Gamma_f){\rm sin}
(2 \omega_k |t-t'|) \Bigr\rbrace {\delta}^3({\bf x}-{\bf x}').
\end{eqnarray}
%
In the limit ${\Gamma}_f \gg 1$(strong coupling limit $ f \gg 1$),
%
\begin{eqnarray}
\langle {\xi}_1(x) {\xi}_1(x') \rangle
& \simeq & \frac{f^2}{\Gamma_f}{\int}\frac{d^3{\bf k}}{(2 \pi)^3}
\frac{1}{4[{\rm cosh}(\beta {\omega}_k)+{\rm cos}
(\beta \Gamma_f)]^2}
\Bigl\lbrace 2{\rm cos}(\beta \Gamma_f)
[{\rm cosh}(\beta {\omega}_k)+{\rm cos}(\beta \Gamma_f) ] \nonumber \\
& - & 2 {\rm sin}(\beta \omega_k){\rm sin}(\beta \Gamma_f){\rm sin}
(2 \omega_k |t-t'|) \Bigr\rbrace
\delta (t-t') {\delta}^3({\bf x}-{\bf x}') \nonumber \\
& \simeq & \frac{f^2}{4{\pi}^2 \Gamma_f} {\int}^{\infty}_0
\frac{d \omega_k \omega_k^2}
{{\rm cosh}(\beta {\omega}_k)+1}
\delta (t-t') {\delta}^3({\bf x}-{\bf x}')  \nonumber \\
& \simeq & 2 \eta_1 T \delta (t-t') {\delta}^3({\bf x}-{\bf x}').
\end{eqnarray}
%
Properly speaking, one cannot take the above limit in the minimal standard
model.
But, we expect that the approximation does not give drastic
changes on the final result as long as we concentrate on a time
scale larger than $\Gamma_f^{-1}$.

Since the ordinary fluctuation-dissipation relation holds as above,
we expect
that the static solution is the  canonical distribution.
Introducing a new variable
%
\begin{equation}
\frac{d{\phi}_c}{dt}= p_{{\phi}_c},
\end{equation}
%
we rewrite eq. (2.31) as
%
\begin{equation}
\frac{dp_{\phi_c}}{dt}=\Delta {\phi}_c-V'_{\rm eff}({\phi}_c)
-\eta_1p_{{\phi}_c} +{\xi}_1.
\end{equation}
%
The Fokker-Planck equation for the above Langevin equation becomes
%
\begin{eqnarray}
\frac{ \partial W[\phi_c(\bfx),p_{\phi_c}(\bfx);t]}{\partial t}&=&
\int d^3x\lnk - \frac{\delta~~}{\delta\phic(\bfx)}
\lmk\frac{\delta\calh}{\delta\pphi(\bfx)}W\rmk
+\frac{\delta~~}{\delta\pphi(\bfx)}
\lkk\lmk\frac{\delta\calh}{\delta\phic(\bfx)}+\eta_1
\frac{\delta\calh}{\delta\pphi(\bfx)}\rmk W\rkk\right.  \nonumber \\
 &~&~~~~~~~~~~~\left.
+T\eta_1 \frac{\delta^2W}{\delta\pphi(\bfx)^2}\rnk \nonumber \\
& = & -\int d^3x \frac{\delta \vec{J}(x)}{\delta \vec{\eta}},
\end{eqnarray}
%
where
%
\[  \vec{\eta}({\bf x})= \left( \begin{array}{c}
                      {\phi}_c(\bfx)     \\
                       p_{{\phi}_c}(\bfx)
                \end{array} \right),\]
%
and $\vec{J}$ is the probability current density with the expression
%
\[  \vec{J}= \left( \begin{array}{cc}
                     0&1      \\
                     -1&-{\eta}_1
                \end{array} \right)
\left(\begin{array}{l}
               \frac{\delta \calh}{\delta {\phi}_c} W
+T\frac{\delta W}{\delta {\phi}_c}   \\
 \frac{\delta \calh}{\delta p_{{\phi}_c}} W
+T\frac{\delta  W}{\delta p_{{\phi}_c}}
                \end{array} \right). \]
%
Here $\calh$ is the Hamiltonian given by
%
\begin{equation}
\calh[{\phi}_c,p_{\phi_c}]:=
{\int}dx^3 \Bigl[\frac{1}{2}\lmk\frac{d {\phi}_c}{dt}\rmk^2
+\frac{1}{2}({\nabla} {\phi}_c)^2+V_{\rm eff}({\phi}_c) \Bigr].
\end{equation}
%
 From the expression of the probability current density we easily find
a static solution of the system as
%
\begin{equation}
W_{\rm st} \propto {\exp {\Bigl[-\frac{{\cal H}[\phi_c,p_{\phi_c}]}
{T}\Bigr]}}.  \label{static}
\end{equation}
%
On the other hand, normalizable dynamical solutions may be expressed in the
form
\cite{FP}
\beq
W[\phic(\bfx),\pphi(\bfx);t]=\sum_n\Psi_n[\phic(\bfx),\pphi(\bfx)]
e^{-\frac{\calh}{2T}}e^{-\Lambda_n t},
\eeq
where the eigenfunction $\Psi_n$ satisfies
\beq
-\Lambda_n\Psi_n=\int d^3x\lkk
-\frac{\delta\calh}{\delta\pphi}\frac{\delta\Psi_n}{\delta\phic}
+\frac{\delta\calh}{\delta\phic}\frac{\delta\Psi_n}{\delta\pphi}
+\frac{\eta_1}{2}\frac{\delta^2\calh}{\delta\pphi^2}\Psi_n
-\frac{\eta_1}{4T}\lmk\frac{\delta\calh}{\delta\pphi}\rmk^2\Psi_n
+\eta_1T\frac{\delta^2\Psi_n}{\delta\pphi^2}\rkk.  \label{seq}
\eeq
The lowest eigen value $\Lambda_0$ is of course zero, while
the next lowest eigen value gives the time scale of relaxation to the
thermal equilibrium.  For its estimation we only have to consider
the Hermitian part of (\ref{seq}) \cite{FP},
\beq
\Lambda^{(H)}_n\Psi^{(H)}_n=\int d^3x  2\eta_1T
\lkk -\frac{1}{2}\frac{\delta^2\Psi^{(H)}_n}{\delta{\pphi}^2(\bfx)}
+\frac{1}{2}\lmk\frac{\pphi(\bfx)}{2T}\rmk^2\Psi^{(H)}_n-
\frac{\delta^{(3)}({\bf 0})}{4T}\Psi^{(H)}_n \rkk.
\eeq
The above equation is an infinite collection of harmonic oscillators
with their ground-state energy subtracted.  We thus find formally
$\Lambda^{(H)}_1=\eta_1$, that is, the time scale of
relaxation is in general given by the inverse of the friction coefficient
in the original
Langevin equation.
The relaxation time reads ${\Delta}t_{\rm relax}
 = {\rm Re}({\Lambda}_1)^{-1} \geq {\Lambda}^{(H)-1}_1 =3{\pi}^{-1}
T^{-1}
\sim m_{\rm ew}^{-1} \sim 10^{-2} {\rm GeV}^{-1}
\sim 10^{-27}s$ for the electroweak Higgs fields.
On the other hand the expansion time scale of the universe in this epoch
is
%
\begin{equation}
{\Delta} t_{{\rm expand}} = \frac{1}{H} \sim 10^{-12}s.
\end{equation}
%
As $\Delta t_{{\rm relax}} \ll \Delta t_{{\rm expand}}$,
the system is almost always in thermal equilibrium.

By integrating over $p_{{\phi}_c}$  in (\ref{static})
we obtain the probability
distribution function $P_{\rm st}[{\phi}_c]$ for ${\phi}_c$:
%
\begin{equation}
P_{\rm st} \propto {\exp {\lkk-\frac{{\cal F}[{\phi}_c]}{T}\rkk}},
\end{equation}
%
where
%
\begin{equation}
{\cal F}[\phi_c]:={\int}d^3x \Bigl[\frac{1}{2}({\nabla} {\phi}_c)^2
+V_{\rm eff}({\phi}_c) \Bigr],
\end{equation}
%
where $V_{\rm eff}(\phi_c)$ is the effective potential with loop
corrections.
Thus, we find the ordinary expression for the probability
distributional
function.

We now obtained the basic
tools for  understanding the dynamics of the electroweak phase transition.
In the next section, we estimate the amplitude the thermal
fluctuation by adopting the Gaussian ansatz for the subcritical
bubbles.


\section{The Dynamics of Subcritical Bubbles}

\subsection{Derivation of the Langevin Equation for the Radius}

Having derived the equilibrium probability distribution function
(2.42), we can calculate the characteristic spatial scale of
thermal fluctuation by adopting, say, a Gaussian ansatz for the
Higgs field as was  done in the previous paper
\cite{shiromizu}. However, since we are interested in the kinematics
of subcritical bubbles, we adopt a variational principle approach to
analyze the effective action directly. We choose a
trial functional of the form
%
\beq
\phi_{\pm}=\phi_0 \exp\lmk-\frac{r^2}{R^2_{\pm} (t)}\rmk,~~~r:=|\bfx|,
\label{eq:ansa}
\eeq
%
to derive the effective Langevin
equation for the radius $R(t)$. The above ansatz is
reasonable because its $O(3)$ symmetry helps to minimize
the free energy and we are treating bubble with a thick wall at or
above the critical temperature.
Here ${\phi}_0$ should be identified with a local minimum of the
effective potential which would become the global minimum below
the critical temperature and is the most expected value apart from
$\phi=0$. We shall investigate the kinematics of subcritical bubbles
with amplitude $\phi_0$. The detail discussion for the above ansatz
has been done in the previous paper\cite{shiromizu}.

Thus the degree of freedom has reduced
to one and we then insert eq. (3.1) into the effective action (2.16).

First we calculate the real part. The result is
%
\begin{eqnarray}
{\rm Re}[{\Gamma}(R_+,R_-)] & = & 2{\pi}^{3/2}{\phi}_0^2{\int}dt
\Bigl[ \frac{15}{32{\sqrt {2}}}(R_+{\dot R}_+^2-R_-{\dot R}_-^2)
-\frac{3}{8{\sqrt {2}}}(R_+-R_-) \nonumber \\
& - & (\frac{1}{8{\sqrt {2}}} m^2 +\frac{{\lambda}^2}{768}{\phi}_0^2)
(R_+^3-R_-^3) \Bigr] \nonumber \\
& - & 2 {\int}dtdt'
{\int}\frac{d^3{\bf k}}{(2 \pi)^3}\Bigl( R_+^3(t)
e^{- \frac{1}{2}k^2R_+^2(t)}-R_-^3(t)
e^{-\frac{1}{2}k^2R_+^2(t)} \Bigr) \nonumber \\
& \times & \Bigl( R_+^3(t')
e^{-\frac{1}{2}k^2R_+^2(t')}-R_-^3(t')
e^{-\frac{1}{2}k^2R_+^2(t')} \Bigr)
{\cal A}({\bf k}, t-t'),
\end{eqnarray}
%
where
%
\begin{equation}
{\cal A}({\bf k}, t-t') :=  f^2\frac{{\pi}^3}{2}{\phi}_0^2
{\int} \frac{d^3{\bf p}}{(2 \pi)^3}{\rm Re}\Bigl[ iS^F_{\alpha \beta}
({\bf p}, t-t')S_F^{\beta \alpha}({\bf p}-{\bf k}, t'-t) \Bigr]
\theta (t-t')
\end{equation}
%
The imaginary part which generates the noise term becomes
%
\begin{eqnarray}
{\rm Im}[{\Gamma}(R_+,R_-)] & = & \frac{1}{2}
{\int}dtdt'{\int}\frac{d^3{\bf k}}{(2 \pi)^3}\Bigl( R_+^3(t)
e^{-\frac{1}{2}k^2R_+^2(t)}-R_-^3(t)
e^{-\frac{1}{2}k^2R_+^2(t)} \Bigr)
\Bigl( R_+^3(t')
e^{-\frac{1}{2}k^2R_+^2(t')}-R_-^3(t')
e^{-\frac{1}{2}k^2R_+^2(t')} \Bigr) \nonumber \\
& \times & {\cal B}({\bf k},t-t'),
\end{eqnarray}
%
where
%
\begin{equation}
{\cal B}({\bf k}, t-t')  :=  f^2\frac{{\pi}^3}{2}{\phi}_0^2
{\int} \frac{d^3{\bf p}}{(2 \pi)^3}{\rm Im}\Bigl[ iS^F_{\alpha \beta}
({\bf p}, t-t')S_F^{\beta \alpha}({\bf p}-{\bf k}, t'-t) \Bigr]
\end{equation}
%
This can be rewritten with an auxiliary fields ${\xi}({\bf k}, t)$:
%
\begin{equation}
{\exp {\lnk i \times i{\rm Im}[{\Gamma}(R_+,R_-)]\rnk}}=
{\int}[d \xi]P[{\xi}]
{\exp {\Bigl[i {\int}\frac{d^3{\bf k}}{(2\pi)^{3/2}}
{\int}dt{\xi}({\bf k},t)
\Bigl(R_+(t)^3e^{-\frac{1}{2}k^2R_+^2(t)}
-R_-(t)^3e^{-\frac{1}{2}k^2R_-^2(t)}\Bigr)\Bigr]}},
\label{eq:gaus}
\end{equation}
%
where
%
\begin{equation}
P[{\xi}]:=N{\exp {\Bigl[-\frac{1}{2}{\int}dtdt'{\int}
\frac{d^3{\bf k}d^3{\bf k}'}{(2 \pi)^3}{\xi}({\bf k}, t){\cal B}^{-1}
({\bf k}, t-t'){\delta}^3({\bf k}-{\bf k}'){\xi}({\bf k}', t')\Bigr]}}.
\end{equation}
%
Thus the effective action for $R_{\pm}(t)$ becomes
%
\begin{equation}
S_{\rm eff}(R_+,R_-):={\rm Re}[\Gamma(R_+,R_-)]
+{\int}\frac{d^3{\bf k}}{(2 \pi)^{3/2}}
{\int}dt{\xi}({\bf k}, t)\Bigl(R_+^3(t)
e^{-\frac{1}{2}k^2R_+^2(t)}
-R_-^3(t)e^{-\frac{1}{2}k^2R_-^2(t)}\Bigr).
\label{eq:eff}
\end{equation}
%
 From ${\delta}S_{\rm eff}/{\delta}R_{\Delta}|_{R_{\Delta}=0}=0$,
we obtain the effective equation for $R_c(t)$:
%
\begin{eqnarray}
\frac{d^2R_c}{dt^2}& + & \frac{1}{2R_c}(\frac{dR_c}{dt})^2+\frac{2}{5}
\frac{1}{R_c}+(\frac{2}{5}m^2+\frac{{\lambda}}{60
{\sqrt {2}}}{\phi}_0^2 + \cdot \cdot \cdot )R_c \nonumber \\
& + &  \frac{16{\sqrt {2}}}{15{\pi}^{3/2}{\phi}_0^2}
{\int}dt'{\int}\frac{d^3{\bf k}}{(2 \pi)^3} {\cal A}
({\bf k}, t-t') \Bigl(3R_c(t)-k^2R_c^3(t)\Bigr)
e^{-\frac{1}{2}k^2R_c^2(t)}R_c^3(t')
e^{-\frac{1}{2}k^2R_c^2(t')}  \nonumber \\
& = & \frac{8{\sqrt {2}}}{15{\pi}^{3/2}{\phi}_0^2}
{\int}\frac{d^3{\bf k}}{(2 \pi)^{3/2}}\Bigl(3R_c(t)
-k^2R_c^3(t)\Bigr)
e^{-\frac{1}{2}k^2R_c^2(t)}{\xi}({\bf k}, t).
\end{eqnarray}
%
Assuming that the time-dependence of $R(t)$ is weak, the last term in
left-hand-side of this equation is expanded as
%
\begin{eqnarray}
F& := &  \frac{16{\sqrt {2}}}{15{\pi}^{3/2}{\phi}_0^2}
{\int}^t_{- \infty} {\int}\frac{d^3{\bf k}}{(2 \pi)^3}
{\cal A}({\bf k}, t-t')\Bigl(3R_c(t)-k^2R_c^3(t)
\Bigr)e^{-\frac{1}{2}k^2R_c^2(t)}R_c^3(t')
e^{-\frac{1}{2}k^2R_c^2(t')} \nonumber \\
& = & \frac{16{\sqrt {2}}}{15{\pi}^{3/2}{\phi}_0^2}
{\int}^t_{- \infty} {\int}\frac{d^3{\bf k}}{(2 \pi)^3}
{\cal A}({\bf k}, t-t')\Bigl(3R_c(t)
-k^2R_c^3(t)
\Bigr)e^{-\frac{1}{2}k^2R_c^2(t)}{R}_c^3(t)
e^{-\frac{1}{2}k^2{R}_c^2(t)} \nonumber \\
& + & \frac{16{\sqrt {2}}}{15{\pi}^{3/2}{\phi}_0^2}R_c^3
{\dot R}_c{\int}^{t}_{- \infty}dt'(t'-t)
{\int}\frac{d^3 {\bf k}}{(2 \pi)^3} {\cal A}
({\bf k}, t-t')\Bigl(3-k^2R_c^2(t)\Bigr)^2
e^{-k^2R_c^2(t)}.
\end{eqnarray}
%
The first and second terms in the last expression give the loop
correction on the effective potential
and friction  term $F_f$, respectively. Consequently $F_f \sim
\eta {\dot R}_c$, where $\eta \sim \eta_1 /4 {\sqrt {2}}$.

Next we turn to the noise term, which we denote as
%
\begin{equation}
{\tilde {\xi}}(t)  :=  \frac{8{\sqrt {2}}}{15{\pi}^{3/2}{\phi}_0^2}
{\int}\frac{d^3{\bf k}}{(2 \pi)^{3/2}}\Bigl(3R_c(t)-k^2
R_c^3(t)\Bigr)
e^{-\frac{1}{2}k^2R_c^2(t)}{\xi}({\bf k}, t).
\end{equation}
%
By using the definition $ \xi ({\bf k}, t) $ the self-correlation
under the limit $\Gamma_f \gg 1$ is given by
%
\begin{equation}
\langle \xi ({\bf k}, t) \xi ({\bf k}', t') \rangle
=  {\cal B}({\bf k}, t-t') {\delta}^3({\bf k}-{\bf k}')
\end{equation}
%
and then the correlation of ${\tilde {\xi}}(t)$ becomes
%
\begin{eqnarray}
\langle {\tilde {\xi}}(t) {\tilde {\xi}}(t') \rangle
& \simeq  & \frac{8}{225{\pi}^3{\phi}_0^2R_c^2}{\int}d^3{\bf k}
{\int}d^3{\bf x} \langle {\xi}_1(x){\xi}_1(x') \rangle
\Bigl(3R_c^2(t)-k^2R_c^4(t) \Bigr)e^{-\frac{1}{2}k^2R_c^2(t)}
\Bigl(3R_c^2(t')-k^2R_c^4(t')\Bigr)e^{-\frac{1}{2}k^2R_c^2(t')}
\nonumber \\
& \simeq & \frac{4{\eta}_1T}{15{\pi}^{3/2}{\phi}_0^2R_c} \delta (t-t')
=:{\tilde {\eta}} \delta (t-t').
\end{eqnarray}
%
In
the above result one might suspect
the $R_c$-dependence of the diffusion coefficient,
but one shall understand by the
later discussion that it is a quite reasonable result.

Consequently the Langevin equation becomes
%
\begin{equation}
\frac{d^2R_c}{dt^2}  +  \frac{1}{2R_c}(\frac{dR_c}{dt})^2+\frac{2}{5}
\frac{1}{R_c}+ \alpha (T) R_c =
-{\eta}\frac{dR_c}{dt} +{\tilde {\xi}}( t),
\end{equation}
%
where $\eta := \frac{\pi}{12{\sqrt {2}} } T$ and
%
\begin{equation}
\alpha (T) =\frac{2}{5}m^2+\frac{{\lambda}}{60{\sqrt {2}}}
{\phi}_0^2+ \cdot \cdot \cdot.
\end{equation}
%

\subsection{The Fokker-Planck Equation for the Radius and its Static
Solution}

In order to find the equilibrium state we transform
Langevin equation eq. (3.14) to the Fokker-Planck equation.
The Langevin equation can be rewritten as
%
\begin{equation}
\frac{dP_c}{dt}  =  \frac{ P_c^2}{2R_c M}-\frac{2}{5}
\frac{M}{ R_c}+ M{\alpha}(T) R_c -{\eta}P_c
+ M {\tilde {\xi}}( t),
\end{equation}
%
where $P_c$ is the canonical momentum defined by
%
\begin{equation}
P_c:=\frac{\partial L_{\rm eff}}{\partial {{\dot R}_c}}=\frac{15{\pi}^{3/2}}
{8{\sqrt {2}}}{\phi}_0^2{\dot R}_cR_c =M{\dot R}_c,
\end{equation}
%
with $ M (T,R_c):=\frac{15{\pi}^{3/2}{\phi}_0^2
R_c}{8{\sqrt {2}}} $ and $L_{\rm eff}$ is the Lagrangian of the deterministic
part
of the Langevin equation:
%
\begin{equation}
L_{\rm eff}(R_c,V_c)=\frac{1}{2 }MV_c^2-\frac{2}{5}M
-\frac{1}{3}M{\alpha}(T)R_c^2.
\end{equation}
%

Now we introduce the  probability distribution function
$W(R_c,P_c;t)$ of $R_c$ and $P_c$. Then we obtain the following Fokker-Planck
equation from the above Langevin equation:
%
\begin{equation}
\frac{\partial}{\partial t}W(R_c, P_c;t)=\frac{\partial}{\partial R_c}
\lmk- \frac{P_c}{M}W\rmk+\frac{\partial}{\partial P_c} \lnk \lkk
- \frac{ P_c^2}{2R_cM}+\frac{2}{5}\frac{M}{R_c}
+M \alpha(T)R_c+\eta P_c\rkk W \rnk
+\frac{1}{2}M^2{\tilde {\eta}}\frac{{\partial}^2}
{\partial P_c^2}W.
\end{equation}
%
The above equation can be expressed by using
the effective Hamiltonian
%
\begin{eqnarray}
H_{\rm eff}(R_c,P_c) & := & P_c{\dot R}_c-L_{\rm eff} \nonumber \\
  & = &  \frac{1}{2M} P_c^2+\frac{2}{5}M
+\frac{1}{3}M{\alpha}(T)R_c^2
\end{eqnarray}
%
as
%
\begin{equation}
\frac{\partial}{\partial t}W(R_c, P_c;t)=\frac{\partial}{\partial R_c}
\Bigl(- \frac{ \partial H_{\rm eff}}{ \partial P_c}W \Bigr)
+\frac{\partial}{\partial P_c} \lkk \lmk\frac{\partial H_{\rm eff}}
{\partial R_c}+ M \eta \frac{\partial H_{\rm eff}}
{\partial P_c} \rmk W \rkk
+\frac{1}{2} M^2{\tilde {\eta}}
\frac{{\partial}^2}{\partial P_c^2}W.
\end{equation}
%
The probability current density becomes
%
\[  \vec{J}= \left( \begin{array}{cc}
                     0 & 1      \\
                     -1 & - M\eta
                \end{array} \right)
\left(\begin{array}{l}
               \frac{\partial H_{\rm eff}}{\partial R_c} {W}
+T\frac{\partial W}{\partial R_c}   \\
 \frac{\partial H_{\rm eff}}{\partial P_c} {W}
+T\frac{\partial W}{\partial P_c}
                \end{array} \right), \]
%
and eq. (3.21) reduces to ${\partial}_tW
+{\nabla} \cdot \vec{J}=0$, where $ \nabla =( {\partial}_{R_c},
{\partial}_{V_c}) $. Here we have used the relation between
the diffusion and friction coefficients:
%
\begin{equation}
\frac{1}{2}M^2{\tilde {\eta}}=TM{\eta}.
\end{equation}
%
Then the static solution becomes
%
\begin{equation}
W_{\rm st}(R,P) \propto {\exp {\lmk-\frac{H_{\rm eff}}{T}\rmk}}.
\end{equation}
%
As in the case of Higgs fields the relaxation time scale
is equal to ${\eta}^{-1}$ and is again much shorter than that of
the expansion of the universe. Thus the distribution is always in the
stationary state.

Finally, we note the fluctuation-dissipation relation exactly holds.
Let us recall the case of the Brownian motion of a massive particle.
In such a system the relation between the diffusion and friction
coefficients is
%
\begin{equation}
({\rm Diffusion ~~coefficient })=({\rm Friction ~~coefficient})
\times ({\rm Temperature}) \times
({\rm Mass}) .
\end{equation}
%
As the mass term corresponds to $ M $ in the present case,
the above relation also holds. The $R_c$-dependence of $W_{\rm st}$
is a direct consequence of the three-dimensional volume effect.
Further this dependence is physically reasonable.
When one compares large bubbles and small bubbles, the
destruction due to the thermal noise is not effective for
large one.

\subsection{The Size of Subcritical Bubbles and the Amplitude of the
Thermal Fluctuation}

Now we can estimate the amplitude of the thermal fluctuation.
We first calculate the averaged radius of the subcritical bubble.
Note that previously all authors have assumed
the correlation length for the typical spatial scale of the thermal
fluctuation. As we will see soon, the scale is a calculable quantity
in high temperature phase.
In this subsection, we extend the above analyses to the
electroweak phase transition. For this purpose we start from
the following effective Hamiltonian corresponding to eq.(3.20):
%
\begin{equation}
H_{\rm eff}(R_c,P_c)=  \frac{1}{2M}P_c^2+\frac{2}{5}M
+\frac{1}{3}M{\alpha}_{\rm ew}(T)R_c^2,
\end{equation}
%
where
%
\begin{equation}
\alpha_{\rm ew} (T) :=
\frac{4}{5}D(T^2-T_2^2)
-\frac{8{\sqrt {2}}}{15{\sqrt {3}}}ET{\phi}_0+\frac{1}{10{\sqrt {2}}}
{\phi}_0^2{\lambda}_T
\end{equation}
%
and
%
\begin{equation}
\phi_0:=\frac{3ET}{2{\lambda}_T}
[1+{\sqrt {1-\frac{8{\lambda}_TD}{9E^2T^2}(T^2-T_2^2)}} ].
\end{equation}
%
In the above expression,
some coefficients can be fixed by electroweak particles:
%
\begin{equation}
D=\frac{1}{24}\Bigl[6(\frac{m_W}{\sigma})^2+3(\frac{m_Z}{\sigma})^2+
6(\frac{m_t}{\sigma})^2 \Bigr] \sim 0.169,
\end{equation}
%
%
\begin{equation}
E=\frac{1}{12 \pi}\Bigl[6(\frac{m_W}{\sigma})^3
+3(\frac{m_Z}{\sigma})^3 \Bigr] \sim 0.00965,
\end{equation}
%
%
\begin{equation}
{\lambda}_T=\lambda- \frac{1}{16{\pi}^2} \Bigl[ {\sum}_B
g_B(\frac{m_B}{\sigma})^4
{\ln {(\frac{m_B^2}{c_BT^2})}} -{\sum}_F g_F(\frac{m_F}{\sigma})^4
{\ln {(\frac{m_F^2}{c_FT^2})}} \Bigr] \sim 0.0350,
\end{equation}
%
%
\begin{equation}
T_2={\sqrt {(m_H^2-8B{\sigma}^2)/4D}},
\end{equation}
%
and
%
\begin{equation}
B=\frac{1}{64{\pi}^2}\Bigl[ 6(\frac{m_W}{\sigma})^4+3
(\frac{m_Z}{\sigma})^4
-12(\frac{m_t}{\sigma})^4  \Bigr] \sim -0.00456,
\end{equation}
%
where we used $m_W=80.6$GeV, $m_Z=91.2$GeV, $m_t=174$GeV and $\sigma = 246 $GeV
\cite{abe}. Further we assumed $m_H=60$GeV which is the minimum value
experimentally allowed.

Let us sketch the potential change  with temperature.  As the
temperature decreases, the
non-symmetric vacuum appears at $T=T_1:=
T_2/{\sqrt {1-9E^2/8{\lambda}_TD}} \sim 93.52$GeV and then the hight
of the two vacua coincides with each other at the
critical temperature $T=T_c:= T_2/{\sqrt {1-E^2/{\lambda}_TD}}
\sim 93.43$GeV. Finally, the barrier between two vacua disappears at
$T=T_2$.

As there exists the interaction with other particles in the
present case, the expression of the friction and diffusion term
obtained in the previous section should be
modified accordingly.
However, as long as the fluctuation dissipation relation holds, which
has been confirmed in the presence of bosonic interaction in
\cite{gl4} and fermionic interaction in the present paper,
 the distribution of Higgs fields is given by
%
\begin{equation}
W_{\rm st} \propto {\exp {\lmk-\frac{H_{\rm eff}}{T}\rmk}}.
\end{equation}
%
Because the system is always in the stationary state, the typical scale
can be calculated by the ordinary canonical averaging.
In the high temperature phase the thermal average of $R$ is given by
%
\begin{eqnarray}
\langle R_c \rangle & := & \frac{{\int}dP_cdR_c R_c{\exp
{[-\frac{H_{\rm eff}}{T}]}}}{{\int}dP_cdR_c
{\exp {[-\frac{H_{\rm eff}}{T}]}}} \nonumber \\
                & = & \frac{{\int}^{\infty}_0dR_cR_c^{3/2}{\exp
                   {[-\frac{2M}{5 T}-\frac{M{\alpha}_{\rm ew}}{3T}R_c^2 ]}}}{
                    {\int}^{\infty}_0dR_cR_c^{1/2}
                    {\exp {[ -\frac{2M}{5T}-\frac{M
                    {\alpha}_{\rm ew}}{3T}R_c^2 ]}}} \nonumber \\
& \sim & \frac{{\int}^{\infty}_0dR_cR_c^{3/2}{\exp {[-\frac{2M}{5T}]}}}
{{\int}^{\infty}_0dR_cR_c^{1/2}
{\exp {[-\frac{2M}{5  T}]}}} \nonumber \\
& = & \frac{2{\sqrt {2}}}{{\pi}^{3/2}}\frac{T}{{\phi}_0^2}
\end{eqnarray}
%
In the low temperature phase, $\alpha_{\rm ew}(T)$ is negative and
accordingly
the  averaged radius diverges. However, this is not relevant. We are
interested in whether
the system has supercooling or not in the {\it high}
temperature phase only.

Having seen that ${\overline R}$ is the important scale of
subcritical bubbles near the critical temperature, we shall now estimate
the amplitude of the fluctuation of $\phi$ around $\phi=0$ on this
particular scale, adopting the trial
function
%
\begin{equation}
\phi={\phi}_A{\exp {\lmk-\frac{r^2}{{\langle R \rangle}^2}\rmk}}.
\end{equation}
%
Then the free energy becomes
%
\begin{equation}
F({\phi}_A, T)=\Bigl[ \frac{3{\pi}^{3/2}}{4{\sqrt {2}}}
{\langle R \rangle}
+\frac{{\pi}^{3/2}}{2{\sqrt {2}}}D(T^2-T^2_2){\langle R \rangle}^3
\Bigr]{\phi}_A^2
-\frac{{\pi}^{3/2}}{3{\sqrt {3}}}ET{\phi}_A^3{\langle R \rangle}^3
+\frac{{\pi}^{3/2}}{32}{\lambda}_T{\phi}_A^4{\langle R \rangle}^3.
\label{eq:freethick}
\end{equation}
%
Thus the RMS amplitude of $\phi$ at the symmetric vacuum is
%
\begin{equation}
{\sqrt {{\overline {{\phi}^2}}}}={\sqrt {\frac{{\int}
d{\phi}_A{\phi}_A^2
e^{-\frac{F({\phi}_A,T)}{T}}}{{\int}d{\phi}_A
e^{-\frac{F({\phi}_A,T)}{T}}}}} \sim
\frac{{\phi}_0}{{\sqrt {3+\frac{16D(T^2-T_2^2)T^2}{{\pi}^3
{\phi}_0^4}}}}.
\end{equation}
%
Its temperature dependence is depicted in Fig.1. In the same way as
in the case of $ \langle R \rangle $, the above approximation result gives
the {\it lower} bound for ${\sqrt {{\overline {{\phi}^2}}}}$.
At $T=T_c$ and $T=T_1$, one obtains numerically
${\sqrt {{\overline {{\phi}^2}}}}(T=T_c)=36.2$GeV and
${\sqrt {{\overline {{\phi}^2}}}}(T=T_1)=28.2$GeV, respectively.
This exceeds the first reflection point ${\phi}_*=\frac{ET}{{\lambda}_T}-
{\sqrt {\frac{E^2T^2}{{\lambda}_T^2}-\frac{2D(T^2-T_2^2)}{3{\lambda}_T}}}$,
which implies that the perturbation theory breaks down and that
one can no longer conclude that electroweak phase transition is
of first order.
%
%

The above argument implies that
the phase transition is not accompanied by
supercooling. The transition rate between two vacua is much
larger than the expansion rate of the universe. In fact, it is
given by
%
\begin{equation}
{\Gamma} \sim m_H{\rm exp}\lmk-\frac{F({\phi}_0,T)}{T}\rmk,
\end{equation}
%
where $F$ is the free energy of subcritical bubble configuration.
As $F \sim T$ during
the high temperature phase\cite{shiromizu},
${\Gamma} \sim 10^2$GeV ( $ \gg H \sim 10^{-13}$GeV). Hence
the fraction of two vacua are determined by the hight of the potential
and therefore the fraction is almost the same at $T=T_c$. This means that
any supercooling does not happen and critical bubbles cannot borne.


\section{Summary and Discussion}

In the present paper we have examined the relaxation of the system from
non-equilibrium states and found
that the system settles down rapidly to the stationary distribution, which
is simply  the ordinary thermal distribution. For the
Higgs fields it is
%
\begin{equation}
P_{\rm st}[{\phi}] \propto {\exp {\lmk-\frac{{\cal F}[{\phi}]}{T}\rmk}},
\end{equation}
%
where ${\cal F}[\phi]$ is the free energy. For the radius of subcritical
bubbles,
%
\begin{equation}
W_{\rm st}(R,P) \propto {\exp {\lmk-\frac{H_{\rm eff}(R,P)}{T}\rmk}},
\end{equation}
%
where $H_{\rm eff}(R,P)$ is the effective Hamiltonian. First we
estimated by using $W_{\rm st}(R,P)$ the mean radius of subcritical
bubbles which determines the spatial scale of the thermal fluctuation.
Next we estimated the field fluctuation by using $P_{\rm st}[{\phi}]$. We
conclude that the electroweak phase transition is quite weakly of
first order and therefore the standard baryogenesis does not work in
minimal standard model, apart from the too small magnitude of CP violation.

Writing down directly the Langevin equation for the radius we have
a glimpse of the kinematics of subcritical bubbles. In the
fluctuation-dissipation relation the dependence on radius is a
natural result from the three volume effect. For larger subcritical
bubble the effect of diffusion is smaller.

Finally we discuss validity of  some approximations employed here.
First in the derivation of the Langevin equation (2.31) we assumed
the approximate homogeneity on the Higgs fields.
In the derivation of the Langevin equation (3.14) for the radius we have
used a similar approximation, which is correct if
$T> \langle R_c {\rangle}^{-1}$. In fact $  \langle R_c {\rangle}^{-1}
/T \sim (E/{\lambda}_T)^2 \sim 0.1 $. Thus the
approximation is justified.
Second,
we have taken the strong coupling limit to obtain the
white noise.  In practice
it is colored and in general the diffusion term is written by a integral
expression, which would be too complicated to treat analytically.  We
expect, however, as long as we focus on a time scale larger than
$\Gamma_f^{-1}$
the more elaborate analysis of the colored noise would yield the same
result and that in this sense our simplified analysis suffices.
We hope to consider this subject  further in a future publication.
Third, we did not work out the calculation of the loop correction
by gauge bosons on the friction and diffusion coefficients. Although
it is essential in order to
obtain the cubic term in the effective potential,
their contribution to the friction
and diffusion coefficients is smaller than top quark.

\vskip 1cm

\centerline{\bf Acknowledgment}
TS thanks H.\ Sato, K.\ Nakao and T.\ Tanaka for discussions.
We would like to a referee for his useful comments.
MM thanks Kurata foundation
for financial support. This work was partly supported by Grant-in-Aid
for Scientific Research Fellowship, No.\ 2925 (TS) and by the
Scientific Research Fund of Ministry of Education, Science,
and Culture, No.\ 06740216 (JY).


%
%
%
%


\begin{center}
{\Large Figure Captions}
\end{center}

\begin{enumerate}

\item{Fig.1:} The RMS of thermal fluctuations. The unit of the
the vertical axis is GeV.

\end{enumerate}

\end{document}